\documentclass[journal=jacsat,manuscript=article]{achemso}

\usepackage[version=3]{mhchem} 

\DeclareMathOperator\erf{erf}
\DeclareMathOperator\sech{sech}

\author{Varvara Zubyuk}
\affiliation{Faculty of Physics, Lomonosov Moscow State University, Moscow 119991, Russia}
\email{zubyuk@nanolab.phys.msu.ru}
\author{Pavel Shafirin}
\affiliation{Faculty of Physics, Lomonosov Moscow State University, Moscow 119991, Russia}
\author{Maxim Shcherbakov}
\affiliation{Faculty of Physics, Lomonosov Moscow State University, Moscow 119991, Russia}
\altaffiliation{School of Applied and Engineering Physics, Ithaca, NY 14853, USA}
\author{Gennady Shvets}
\altaffiliation{School of Applied and Engineering Physics, Ithaca, NY 14853, USA}
\author{Andrey Fedyanin}
\affiliation{Faculty of Physics, Lomonosov Moscow State University, Moscow 119991, Russia}

\title
  {Externally driven nonlinear time-variant  metasurfaces
}

\abbreviations{IR,NMR,UV}
\keywords{Metasurfaces, nonlinear optics, time-variant materials, ultrafast processes}

\begin{document}




\begin{abstract}
Resonant photonic nanostructures exhibiting enhanced nonlinear responses and efficient frequency conversion are an emergent platform in nonlinear optics. High-index semiconductor metasurfaces with rapidly tuned high-Q resonances enable a novel class of time-variant metasurfaces, which expands the toolbox of color management at the nanoscale. Here, we report on the dynamic control of the nonlinear optical response in time-variant semiconductor metasurfaces supporting high-quality factor resonances in the near-infrared spectral range. Pump--probe measurements of germanium metasurfaces at negative pump--probe time delays reveals frequency conversion in the fundamental beam and a blue-shift of 10~nm ({3.05$\omega$}) and 40\% broadening in the third harmonic signal  due to the photoinduced time-variant refractive index. A time-dependent coupled-mode theory, in excellent agreement with the experimental data, validated the time-variant nature of the system. Our findings expand the scope of time-variant metasurfaces and may serve as base for the next generation of nanoscale pulse shapers, optical switches and light sources.


\end{abstract}

\section{Introduction}

Metasurfaces~--- arrays of periodically arranged  nanoparticles ~\cite{kuznetsov2016optically,yu2014flat} --- have established themselves as a promising alternative for various optics elements like lenses\cite{khorasaninejad2017metalenses}, holograms\cite{ni2013metasurface, zheng2015metasurface}, deflectors\cite{wang2018nonlinear, zhou2017efficient} and polarizers \cite{yu2014flat}. As an emergent platform in nonlinear optics, resonant photonic nanostructures demonstrated enhanced nonlinear responses and efficient frequency conversion \cite{shcherbakov2014enhanced, makarov2015tuning, grinblat2016enhanced, shibanuma2017efficient,wang2018nonlinear, semmlinger2018vacuum, liu2018all, liu2018enhanced,sain2019nonlinear}. One of the downsides of metasurfaces is their fixed optical properties, imposing restrictions on their use after fabrication. Many works were dedicated to active and tunable metasurfaces, where an external stimulus can change the properties of the structures. As such,  mechanical~\cite{holsteen2017purcell}, electrical~\cite{park2017dynamic} or thermal control~\cite{rahmani2017reversible}, optical excitation\cite{yang2015nonlinear, shcherbakov2015ultrafast, shcherbakov2017ultrafast}, magneto-optical control\cite{zubritskaya2018magnetic}, use of phase-change materials\cite{gholipour2013all, karvounis2016all} or chemical approaches\cite{di2016nanoscale} and hybrid systems like liquid-crystals\cite{sautter2015active, komar2017electrically} can tune the spectral position of metasurfaces' resonances. 
Many of the potential applications of metasurfaces, especially in nonlinear and tunable cases, require sharp spectral features and high local electric fields. 
To fulfill these conditions, Mie-type resonances are widely used\cite{kruk2017functional}, first demonstrated in spherical high-index particles and other geometries such as disc-shaped particles~\cite{kuznetsov2016optically}. Single resonant nanoparticles can be combined into strongly coupled arrays, where high quality factor (high-Q) modes enable metasurfaces with sharp spectral features and enhanced light-matter interactions \cite{neuner2013efficient, sui2018high}, leading to many exciting applications \cite{shalaev2015high, liu2017optical}. Using high-Q structures, photonic devices with properties beyond what is available with bulk materials and low-Q metasurfaces can be attained \cite{reshef2019multiresonant, zubyuk2019low, shcherbakov2019photon}. High-index semiconductor metasurfaces with specially designed high-Q resonances have great capability for enhanced nonlinear effects and other benefits for creating various compact nonlinear photonics devices.


In the extreme case where resonances of metasurfaces are tuned rapidly, a novel class of effects pertaining to the time-variant nature of metasurfaces emerges. Under this framework, harmonic waves cease to be solutions of Maxwell's equations, and effects such as frequency conversion~\cite{lee2018linear, lee2019electrical, karl2020frequency}, photon acceleration~\cite{shcherbakov2019photon}, nonreciprocal light reflection, beam steering or focusing~\cite{guo2019nonreciprocal, zang2019nonreciprocal, cardin2020surface} that are impossible in stationary systems rise, similar to original works in rapidly generated plasmas~\cite{kuo1990frequency, dias1997experimental}. The frequency conversion that has recently been demonstrated for metasurfaces also has origins similar to that of the coherent artifact in ultrafast time-resolved infrared (IR) spectroscopy of vibrational transitions in chemical systems~\cite{hamm1995coherent, yan2011perturbed} and for semiconductors~\cite{fluegel1987femtosecond, joffre1988coherent}, which manifest at negative time delays, when the probe pulse precedes the pump pulse. Even though many systems have demonstrated linear frequency conversion due to abrupt time modulation, ultrafast studies of nonlinear responses \cite{utikal2010all, shorokhov2016ultrafast,sartorello2016ultrafast,cheng2020ultrafast}, which are prominent in resonant metasurfaces, are yet to unveil the control of light frequency for nonlinearly generated waves.


Here, we report on the dynamic control of the nonlinear optical response of a time-variant semiconductor metasurface. We design and experimentally implement a germanium-based metasurface that possesses a high-Q resonance in the near-infrared spectral range and shows frequency conversion due to the rapidly photoinduced time-variant refractive index in a femtosecond pump--probe experiment. 
Owing to the combination of the high-Q resonance and time-variant refractive index, the metasurface demonstrates blue-shifted and spectrally broadened fundamental beam and third harmonic generation at negative pump--probe time delays.
We observe a  blue-shift of 10~nm and 40\% broadening in the third harmonic generation spectrum around a wavelength of 540~nm, as well as its complete suppression at positive delays.
To model the experimental results, a coupled-mode theory is developed, which has confirmed the observed wavelength shifts and validated their time-variant nature. The observed external photoinduced control over the frequency of nonlinearly generated photons establishes a new utility for time-variant metasurfaces and opens opportunities for novel light sources at the nanoscale.

\section{Results and discussion}
\subsection{Metasurface samples and concept.} 

\begin{figure*}
\includegraphics[width=0.75\paperwidth]{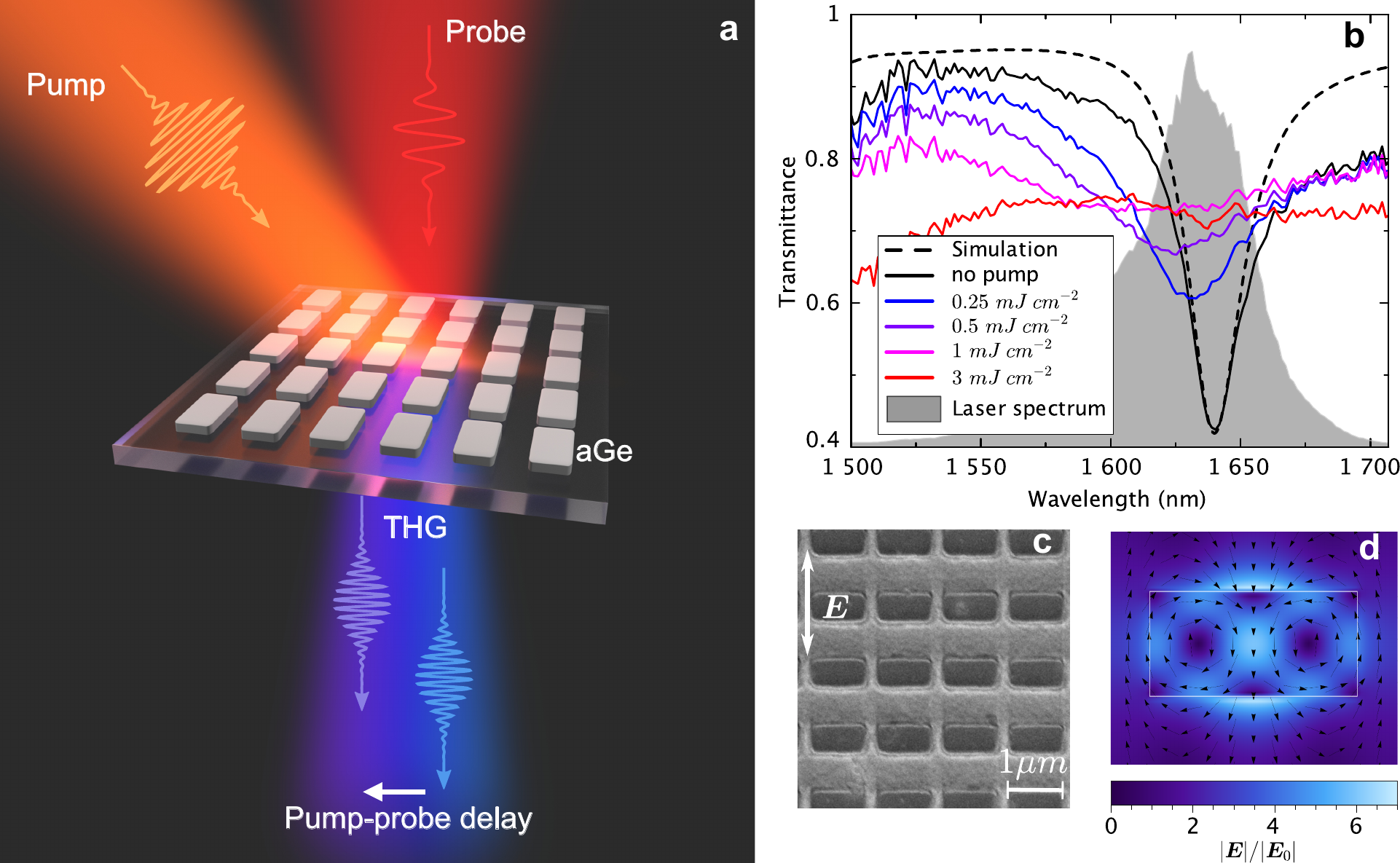}
\caption{\label{fig:1} \textbf{External optical drive enables time-variant resonances in a nonlinear germanium metasurface.} \textbf{a}, The concept of optically driven nonlinear response of a time-variant amorphous germanium (aGe) metasurface. The absorption of a pump pulse rapidly modifies the refractive index of aGe. As a result, the third harmonic signal generated by the probe beam can get dynamically up-converted, following the rapid evolution of the resonance. \textbf{b}, The experimental and simulated transmittance spectra of the aGe metasurface in the absence of optical pumping (solid and dashed black curves, respectively) and transmittance spectra after illumination by different pump fluences (colored curves). The grey region shows the spectrum of the probe pulse. \textbf{c}, A scanning electron microscope image of a typical metasurface sample. The white arrow represents the polarization direction of the probe pulse. \textbf{d}, A map of the simulated field amplitude $|\textbf{E}/\textbf{E}_0|$ (color) and direction (black arrows) at the central wavelength of the metasurface resonance inside  the aGe cuboid showing high local field enhancement pertaining to the high-Q resonance.}
\end{figure*}

This work is conceptualized in Fig.~\ref{fig:1}a. A semicondictor metasurface formed by an array of amorphous Ge (aGe) cuboids was chosen as a time-variant medium, and its capabilities as a frequency converter were investigated using pump--probe spectroscopy. Due to aGe cuboids lying close to each other, the metasurfaces showed high-Q resonances\cite{shalaev2015high}, and aGe provided the free-carriers generation under optical pump illumination and, therefore, enabled rapid changes of the refractive index; see Methods for the sample fabrication procedure. A scanning electron microscope (SEM) image of a typical metasurface is shown in Fig.~\ref{fig:1}c. The Q-factor of the resonances depended on the size of the gap between the short sides of the cuboids and varied from 30 to 65. We chose the metasurface with $Q=65$ as a sample for this study. Its experimental and simulated transmittance spectra in the near-IR in unperturbed regime are shown in Fig.~\ref{fig:1}b by solid and dashed black curves, respectively; see Methods for the calculation and measurement details. The polarization of the incident light was perpendicular to the long sides of the cuboids, shown on the SEM image with a white arrow. The grey region shows the spectrum of the incident probe pulse, which resonantly excited a magnetic quadrupolar mode in the metasurface. In Fig.1d, a calculated field distribution inside the aGe cuboid at the resonant wavelength shows two electric field vortices that generate out-of-phase, out-of-plane magnetic dipoles. The transmittance spectra after illumination by different pump fluence (colored curves) exhibit a shift of the resonance and a decrease in its Q-factor due to the modification in the aGe refractive properties. These changes increase with increasing the pump fluence, as more free-carriers are generated, and for a pump fluence of 1~$\text{mJ/cm}^{2}$ (pink curve), the resonance is still visible and experiences a blue-shift of about of 30~nm. We estimate the resonance shift at a fluence of 3~$\text{mJ/cm}^2$ to be up to approximately 100~nm.

\subsection{Transient linear and nonlinear spectroscopy.}

The ultrafast dynamics of infrared transmittance and third harmonic generation (THG) signal were studied using the pump--probe technique. The setup is schematically shown in Fig.~\ref{fig:2}a. Ti:Sapphire amplified femtosecond pulses were used to pump the metasurface and induce rapid changes of the refractive index. The pulses from the parametric amplifier were used to probe the response of metasurfaces. The system ran at a low repetition rate of 1~kHz to minimize the residual pulse-to-pulse heating. The probe pulse length was 50~fs and its central wavelength was tuned to the metasurface resonance. The pump pulse was 60-fs-long, had a wavelength of 800~nm and controllable varied fluence from 0.25 to 3~$\text{mJ/cm}^{2}$. Both beams were $p$-polarized at the metasurface plane. The probe was focused under normal incidence, whereas the pump fell on the sample at a small angle from the normal (for more details, see Methods). The transmitted probe pulse spectra were measured under and without the pump presence as a function of both time delay $\tau$ between the pump and probe pulses in the near-infrared and visible spectrale ranges. The transient transmittance is defined as: 
\\$\Delta T/T~=~(T_{\rm{pump}} - T)/T~=~\Delta I/I~=~(I_{\rm{pump}} - I)/I $, where $T$ is the transmittance of the sample and $I$ is the laser spectrum of the probe passing through the sample in the absence of the pump pulse, and $T_{\rm{pump}}$, $I_{\rm{pump}}$ are the values under the pump illumination.

\subsection{Frequency conversion in fundamental spectrum.}

\begin{figure*}
\includegraphics[width=0.95\textwidth]{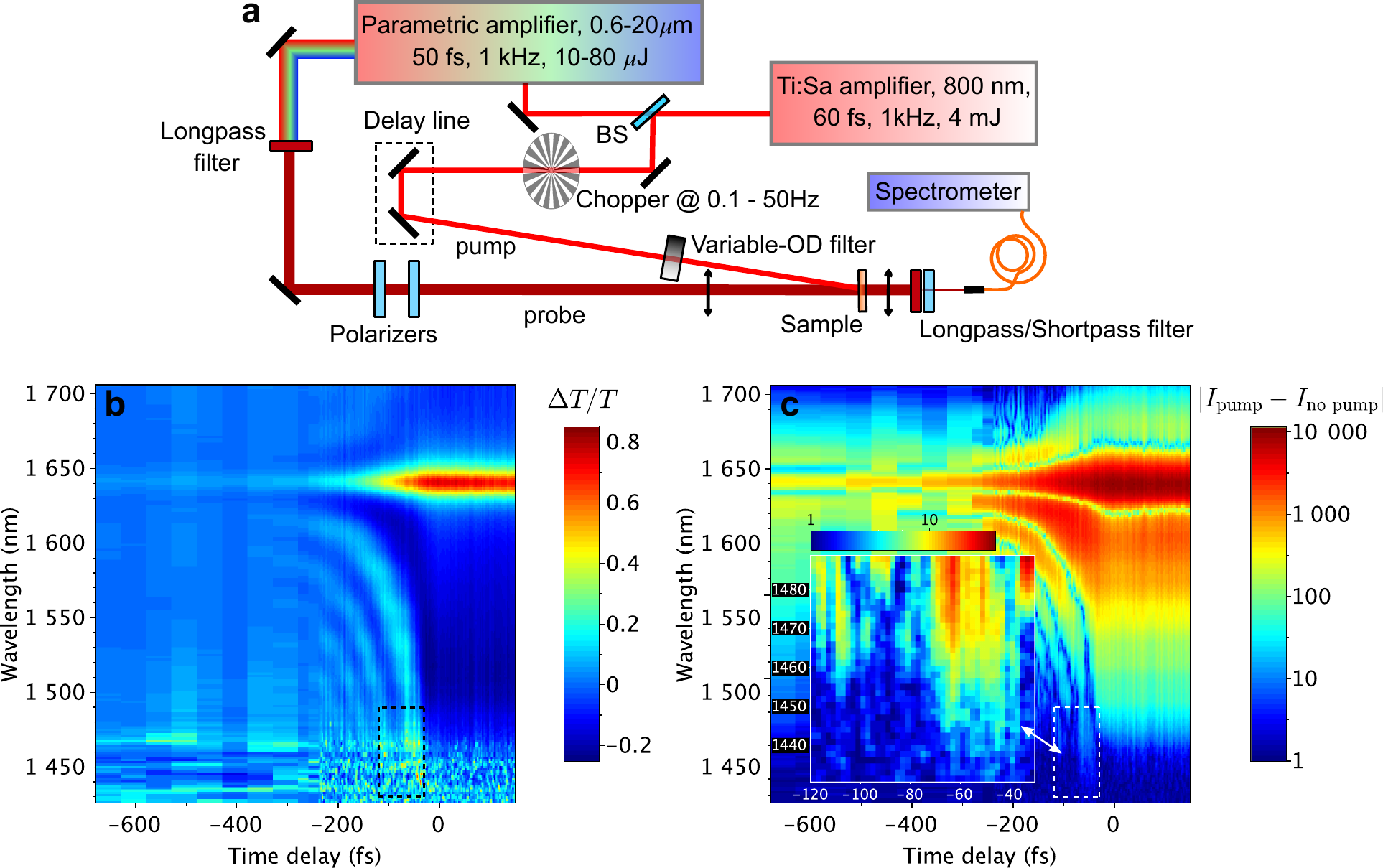}
\caption{\label{fig:2} \textbf{Experimental linear frequency conversion in a time-variant aGe metasurface.} \textbf{a}, Pump--probe spectroscopy setup. \textbf{b}, A 3D-map of transient transmittance for a metasurface with the resonance at $\lambda_R = $~1640 nm. The dashed black rectangle denotes the zoom area for insert in panel \textbf{c}. \textbf{c}, The 3D-map of the difference between probe spectrum under pump illumination and without pump illumination, logarithmic scale. Inset: zoom of the dashed white rectangle area showing new frequency components generated at wavelenghts as short as 1440~--~1450~nm. }
\end{figure*}

Photoinduced modulation of the metasurface's resonance enables frequency conversion in the fundamental beam, as seen in Fig.~\ref{fig:2}b,c. Figure~\ref{fig:2}b demonstrates transient transmittance $\Delta T/T$ for an aGe metasurface with the resonance at the central wavelength $\lambda_R$ = 1640~nm. A significant change in transmission is observed, both positive (up to +80\%) and negative variation (down to $-25$\%). At negative time delays, characteristic fringes~\cite{yan2011perturbed, shcherbakov2019time, karl2020frequency} can be observed, 
originating from the interference between the resonant part of the probe that is being dynamically frequency-converted in the metasurface and the non-resonant part of the probe. The extent to which the new frequencies are generated in the metasurface through time variance can be appreciated at wavelengths far from both the unperturbed resonance and the probe carrier wavelength. 
The differential intensity plot in Fig.~\ref{fig:2}c shows that at small negative pump--probe delays around $-70$ to $-40$~fs, above-noise signal can be generated at wavelengths as short as 1440~--~1450~nm, which is about 200~nm away from the initial resonance position, signifying strong frequency conversion due to the time-variant nature of the metasurface.

\subsection{Frequency conversion in the third harmonic signal.}

\begin{figure*}
\includegraphics[width=0.7\paperwidth]{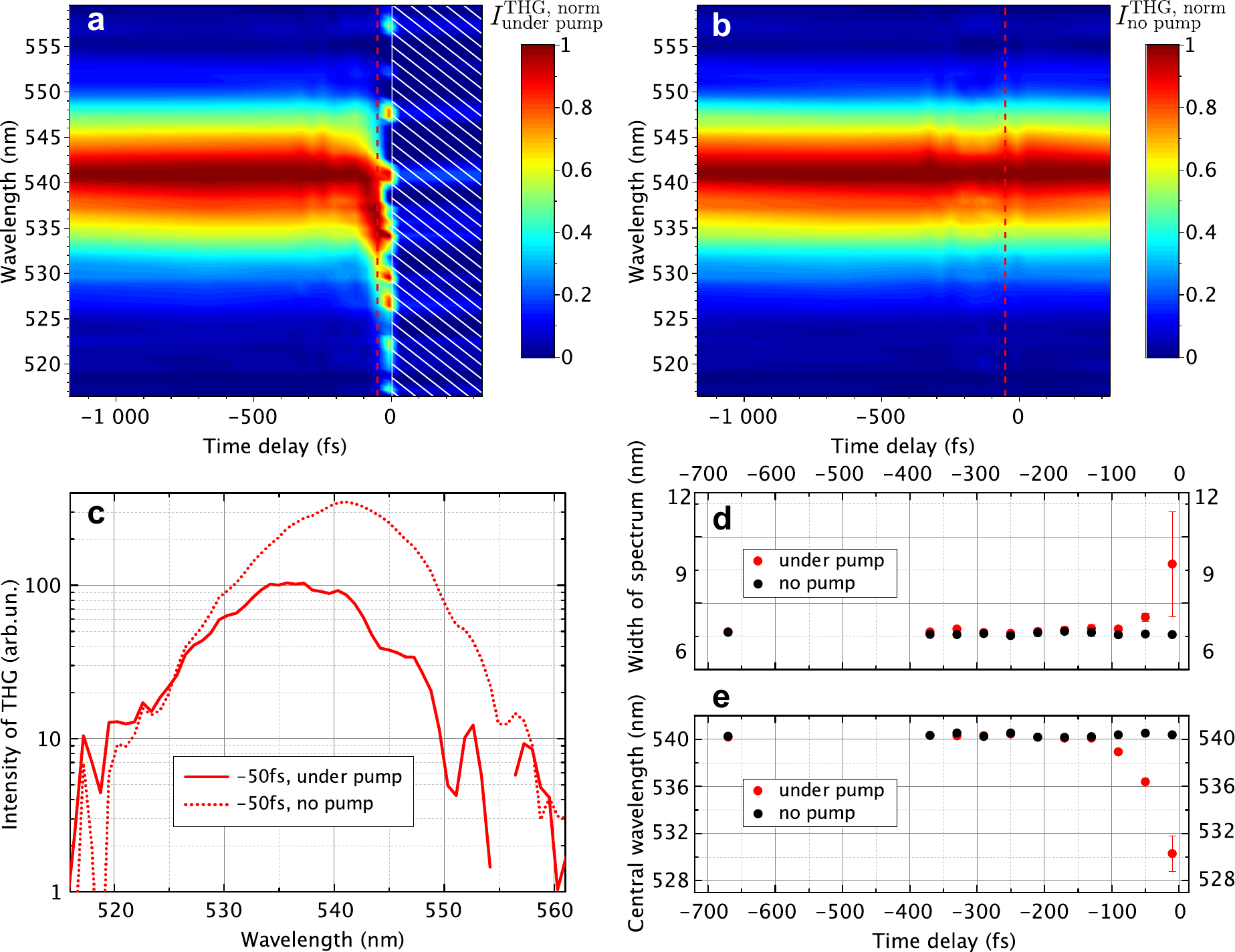}
\caption{\label{fig:3} {\bf Frequency conversion of the harmonic response in a time-variant metasurface.} The normalized spectra of the third harmonic signal generated  by the probe pulse under pump (\textbf{a}) and without pump illumination (\textbf{b}) as a function of the pump--probe delay. Red dashed lines in \textbf{a} denote the time delays for the THG intensity spectra cross-sections in \textbf{c}. The hatched white rectangle marks the area with noisy signal that prevents meaningful normalization. \textbf{c}, THG spectra at delay $\tau = - 50$~fs under pump illumination (solid curve) and without pump illumination (dotted curve). \textbf{d}, The extracted width $\sigma$ and \textbf{e}, central wavelength $\lambda_0$ of the THG spectrum as functions of $\tau$ under pump illumination (red dots) and without pump illumination (black dots). }
\end{figure*}

Due to the high values of the intrinsic third-order nonlinear susceptibility $\chi^{(3)}$ of aGe, one can observe the third harmonic generation generated by the probe pulse in the aGe material. The THG is described by the nonlinear polarization $\textbf{P}^{(3)}_{NL} (3\omega) = \varepsilon_0 \chi^{(3)}(3\omega)\vdots\textbf{E}(\omega)\textbf{E}(\omega)\textbf{E}(\omega)$, and, owing to the electric field enhancement in the metasurface, the THG signal is observed at a lower power compared to the bulk aGe. Here, we show how the free-carrier-induced time-dependent refractive index of the metasurface imposes femtosecond dynamics in the THG response, causing conversion of the THG frequency. Experimental ultrafast all-optical nonlinear modulation of aGe metasurfaces is shown in Fig.~\ref{fig:3}a. The normalized third harmonic signal generated by the probe pulse under pump illumination demonstrates blue-shift of the THG wavelength at the femtosecond scale for negative pump--probe time delays. This blue-shift (or photon acceleration~\cite{dias1997experimental}) of nonlinear signal occurs due to the negative term of the photodoped semiconductor refractive index  that is proportional to the number of free carriers generated by the pump beam $\Delta n \propto -N_{FC}(t)$. The hatched white rectangle marks the area with noisy signal and with corresponding artificially low values that prevents meaningful normalization. The unperturbed aGe metasurface generates third harmonic signal with stable behaviour in time (Fig.~\ref{fig:3}b). Red dashed lines in Fig.~\ref{fig:3}a,b denote the time delay for the THG intensity spectra cross-sections in Fig.~\ref{fig:3}c. 

Figure~\ref{fig:3}c shows unnormalized individual THG spectra at time delay value $\tau = - 50$~fs under pump illumination (solid curve) and without pump illumination (dotted curve). The THG signal demonstrates blue-shifting and broadening whose values increase with increasing the time delay. We fit the spectra in Fig.~\ref{fig:3}a,b by a $\sech^2$ function:

 \begin{equation}
y = y_0 + A\sech^2\left(\frac{\lambda-\lambda_0}{\sigma}\right),
\label{sech}
\end{equation}
where $y_0$ is the offset, $A$~--- amplitude, $\lambda_0$~--- central wavelength and $\sigma$ is the spectral full width at half maximum. The extracted values of $\sigma$ and $\lambda_0$ are shown in Fig.~\ref{fig:3}d,e, respectively, as a function of $\tau$ under pump illumination (red dots) and without pump illumination (black dots). The highest values of measured blue-shift and broadening of the THG spectrum are 10~nm and 40\%, respectively, found at time delay $\tau=-10$~fs. The blue-shift of 10~nm corresponds to the frequency conversion of 3.05$\omega$ in comparison with unperturbed nonlinear signal that is expected at 3$\omega$. 
The nature of the THG blue-shift effect is associated with the metasurface resonance blue-shift at the fundamental wavelength under pump illumination and subsequent redistribution of the frequency components of the third harmonic spectrum. As shown in Fig.~\ref{fig:1}b, the higher the pump fluence, the larger the blue-shift of the resonance. The nonlinear modulation of the aGe metasurface was measured at a pump fluence of 3~$\text{mJ/cm}^{2}$ when the resonance shift is estimated to be about 100~nm. The basic dynamics related with the shift of the resonance at fundamental frequency occurs at the time scale of $\tau\sim-100$~fs. This region is characterized by the most pronounced and spectrally broad fringes (Fig.~\ref{fig:2}b,c at $\tau$  between $-200$ and $0$~fs). Similar temporal dynamics one can observe in the nonlinear response. 
Finally, when the resonance shifts far enough from the fundamental spectrum center, near $\tau=0$ the nonlinear response becomes highly impeded by the formed plasma; the signal for $\tau>0$ can no longer be detected under the same experimental conditions (hatched white rectangle area in Fig.~\ref{fig:3}a).

\subsection{Theory}
\begin{figure*}
\includegraphics[width=0.75\paperwidth]{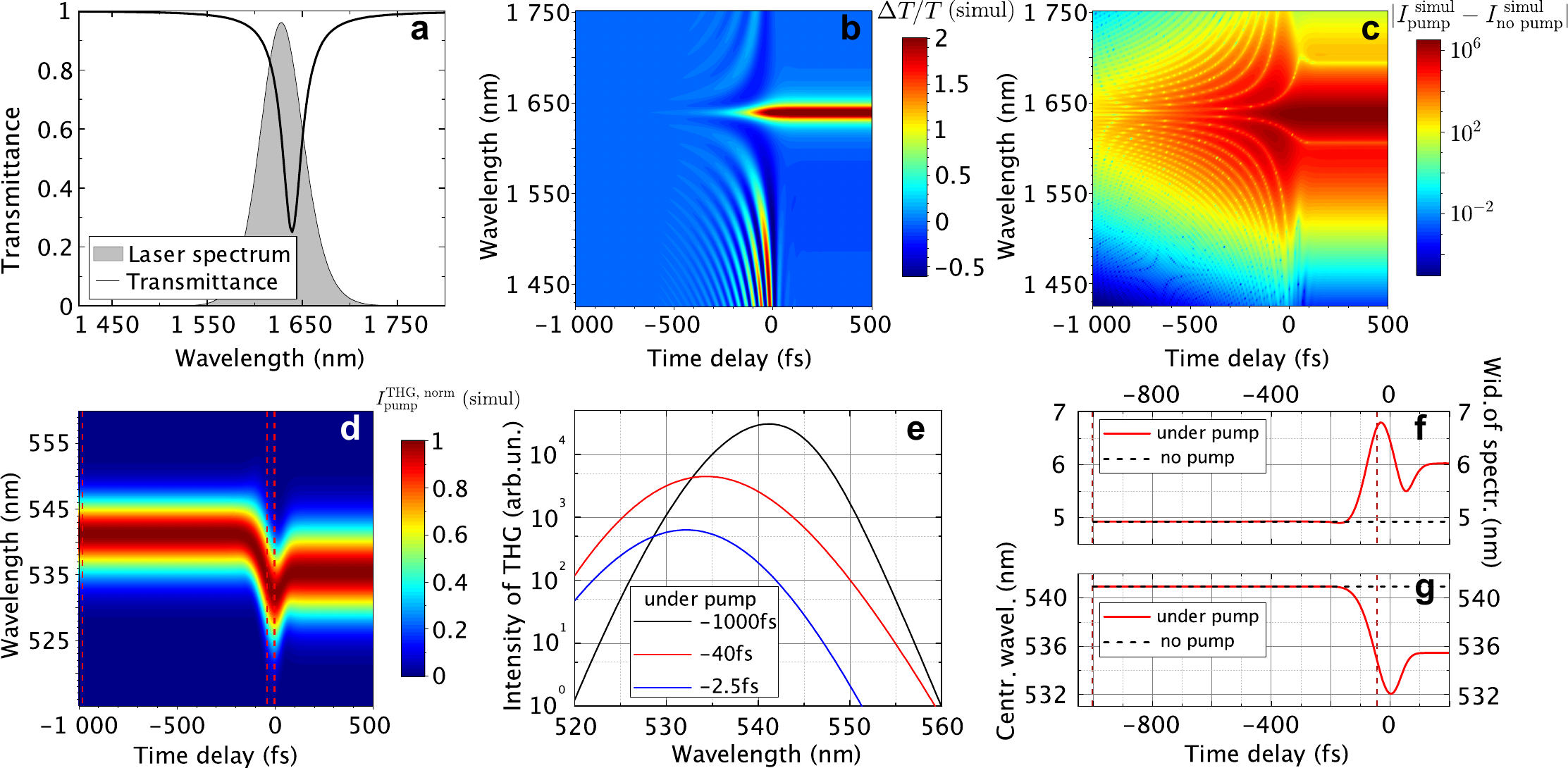}
\caption{\label{fig:4} {\bf  Coupled-mode theory results.} \textbf{a}, Simulated transmittance spectra of the metasurface with the resonance at $\lambda_R = $~1640~nm in the absence of optical pumping (black curve) and the spectrum of the probe pulse (gray region). \textbf{b}, A 3D-map of transient transmittance for the metasurface as a function of wavelength and pump--probe time delay. \textbf{c}, The 3D-map of the difference between probe spectrum under pump illumination and without pump illumination as a function of wavelength and pump--probe time delay. \textbf{d}, The normalized third harmonic signal generated by the probe pulse under pump illumination. Red dashed lines denote the time delays for the THG intensity spectra cross-sections at \textbf{e}. \textbf{e}, THG spectra under pump illumination at $\tau = - 1000, - 40$ and $-2.5$~fs of pump--probe delay (here $I^{\rm THG,~pump}_{\rm -1000 fs} = I^{\rm THG,~no~pump}_{\rm -1000 fs}$). \textbf{f}, The fitting result of THG intensity by $\mbox{sech}^2$ function, Eq.(\ref{sech}), with spectral width $\sigma$ and central wavelength $\lambda_0$ (\textbf{g}) under pump (red curves) and without pump illumination (dashed black curves) as a function of time delay between the pump--probe pulses. Red dashed lines denote the time delays $\tau = - 1000$ and $-40$~fs for cross-sections at \textbf{e}.}
\end{figure*}

To confirm the description of our experiment and to get a better understanding of the system dynamics, we construct a time-dependent coupled-mode theory (CMT)~\cite{haus1984waves}. This theory allows to simulate the ultrafast all-optical modulation of the metasurface's resonant mode as a function of both time delay between the pump--probe pulses and probe wavelength. The probe is the short $\sech^2$ pulse (Fig.~\ref{fig:4}a, grey region) and the pump influence is imitated by the time dependence of the mode parameters: central frequency $\omega_0$ and damping factor $\gamma$. The transmittance of the unperturbed mode is shown in Fig.~\ref{fig:4}a and is quite similar to the transmittance of the experimental aGe metasurface (Fig.~\ref{fig:1}b black curve). The probe spectrum width $\sigma = 50$~nm is also taken close to the experimental value of $\sigma = 55$~nm. Figure~\ref{fig:4}b demonstrates the calculated transient transmittance of the metasurface, which has excellent agreement with the experiment. The region with high modulation values (dark red area at short wavelengths) occurs due to the division by small values of the unperturbed spectrum, which may not describe well the experimental probe spectrum as far as the spectral tails are concerned. 
This area can be clearly seen without division from 3D-map of the absolute value of difference between probe spectrum under pump illumination and without pump illumination on a logarithmic scale (Fig.~\ref{fig:4}c). 
The corresponding normalized third harmonic signal is given in Fig.~\ref{fig:4}d. The signal demonstrates the blue-shift at the femtosecond scale as obtained in the experiment. The red dashed lines denote the time delays for the intensity spectra cross-sections in Fig.~\ref{fig:4}e. These cross-sections of THG spectra under pump illumination display the blue-shifting and broadening of the spectra which increase at $\tau\approx0$. The THG probe spectrum at $\tau = - 1000$~fs under pump (black curve) completely matches with the THG probe spectrum without pump in this region. The temporal dynamics for width of spectrum $\sigma$ and central wavelength $\lambda_0$ was obtained from fitting by $\sech^2$ function with Eq.(\ref{sech}) as a function of time delay between the pump--probe pulses (Fig.~\ref{fig:4}f,g). The blue-shift reaches a maximum value of about 10~nm and the broadening is about 40\%, which is in excellent agreement with the experimental values.

Due to the high-Q factor of the metasurface's mode, a part of the probe pulse that couples to the metasurface stays with it for a time comparable to the lifetime of the mode. The powerful pump pulse arrives shortly after the mode is populated and generates free-carriers, altering the refractive index of the semiconductor and shifting the resonance by up to 100~nm. This shift, along with the dynamic widening of the resonance, enables broadband frequency conversion at up to 200~nm away from the central wavelength at negative time delays in pump--probe traces. The femtosecond-scale changes in the fundamental beam spectrum lead to the femtosecond changes in the third harmonic signal generated by the probe pulse and manifested in the blue-shift and broadening of the third harmonic spectrum. While this observation is deemed possible in high-Q metasurfaces, our modelling reveals that the femtosecond dynamics of THG at negative time delays $\tau<0$ demonstrated in Fig.~\ref{fig:3} is significantly smaller for lower Q-factors, see Supplementary Fig. S1. Our results pinpoint the importance of high-Q resonances in the processes of frequency conversion in time-variant metasurfaces and serve as an important step to the  full understanding of this phenomenon.

To conclude, we have designed, fabricated and studied an aGe-based metasurface that shows photoinduced frequency conversion via a resonance with rapidly varying parameters. Pump--probe experiments show pronounced spectral features in the probe beam and its third optical harmonic at negative pump--probe delay times. In the first observation of switching and frequency tuning of a harmonic signal in a semiconductor metasurface, we demonstrate a 10~nm blue-shift and 40\% broadening of its spectrum at negative pump--probe delays and almost complete suppression of the THG at positive delays. The observed blue-shift and broadening are caused by the pump-induced dynamics of the high-Q factor metasurface resonance and due to high light-matter interaction observed at low energy, as confirmed by coupled-mode theory calculations. Nonlinear time-variant metasurfaces represent a promising novel class of artificial semiconductor nanostructures that can serve as base for new-generation pulse shapers, optical switches and light sources.

\footnotesize{
\section{Methods}
\subsection{Sample fabrication.}
Samples of germanium metasurfaces were fabricated by thermal deposition of a thin (200~nm) aGe film on a CaF$_2$ substrate, subject to electron beam lithography with 100-nm-thick HSQ resist exposed at a dose of 800~$\mu$C/cm$^2$, development in TMAH (120~s, 3\% solution), and reactive ion etching in SF$_6$ plasma. The total dimensions of the nanostructured areas were 500 by 500~$\mu \text{m}^2$.

\subsection{Sample Characterization.}
Near-infrared spectroscopy of the samples was performed using a home-built setup. The beam from an incandescent light source was focused to a spot size of about 300~$\mu\text{m}$ with a numerical aperture of NA = 0.05. The transmitted beam was analyzed by an IR InGaAs CCD-based  spectrometer. Normalization was done by the spectrum of the source without sample. The setup allowed to measure transmittance for different angles of incidence and linear polarization states of the incident light. The resonances with  high Q-factors were observed for the  polarization orientation of the incident light perpendicular to the long sides of the aGe cuboids. 

\subsection{Pump–-probe spectroscopy.}
A Ti:Sapphire regenerative amplifier serving as a pump, and an optical parametric amplifier (OPA) serving as a probe were used to investigate frequency conversion in the semiconductor metasurface. The pump was a 60~fs laser pulse train (as measured using an autocorrelator) of 800~nm wavelength with 1~kHz repetition rate. Its average power was varied using a variable optical density filter from 0.5 to 6~mW, while retaining a spot size of about 500~$\mu$m in diameter in the sample plane. The probe was a 50~fs pulse train with 1~kHz repetition rate and (idler) wavelength tunable in the range from 1580 to 2600~nm, with an average beam power of about 40~mW. The duration of the probe was measured by cross-correlation scheme with the use of pump as one of the pulses using a BBO nonlinear crystal that produced a sum-frequency beam. The transmitted probe pulse spectra (third harmonic generated by the probe) were measured as a function of both time delay between the pump and probe pulses and the probe fundamental wavelength (THG wavelength) under and without pump presence by a near-IR spectrometer (visible spectrometer). The measurements were carried out with a frequency from 0.1 to 50~Hz depending on the averaging time of the spectrometer. A normalized 3D map of THG intensity was obtained by normalizing the spectrum at each pump--probe time delay to its maximum. The normalized intensity of THG was used for a clearer observation of the signal blue-shift due to the decreasing of the signal value with increasing time delay of pump--probe.

\subsection{FDTD simulations}
The transmittance spectra were calculated using a commercial Lumerical FDTD Solutions software for an aGe metasurface with the following parameters: the dimensions of the cuboid short side $x=360$~nm, long side $y=753$~nm, the period along the \textit{Ox} axis $P_x=1035$~nm, along the \textit{Oy} axis $P_y=1060$~nm and the aGe layer thickness of 140~nm. The characteristic geometric parameters was achieved from the SEM image of the sample. There was a residual silica cap due to the fabrication process on top of the aGe cuboid. We used a non-dispersive dielectric material with $n = 1.6$ and $h_{\rm SiO_x} = 100$~nm for this cap and $n = 1.45$ for the CaF$_2$ substrate. For the aGe material, we used the available tabulated experimental data~\cite{palik1998handbook}. Periodic boundary conditions were used to simulate periodic structure of metasurface and perfectly matched layer (PML) conditions were used in $z$ direction. The plane wave excitation was from the top at a normal angle with the polarization perpendicular to the long sides of the cuboids, as shown by a white arrow in an SEM image of the sample in Fig.~\ref{fig:1}c.

\subsection{CMT calculations.}

The experimental data can be understood in terms of a time-dependent coupled-mode theory. We considered a single resonant mode being excited by a short pulse, for which CMT gives the following equation:

\begin{equation}
\dot{a}(t) + [ i\omega(t) + \gamma_{nr}(t) + \gamma_{r}]a(t) = \sqrt{\gamma_{r}(t)}s(t)
\label{"CMT"}
\end{equation}

Here, $a(t)$ is the complex amplitude of the exited mode, $\omega$ is the resonant frequency of the mode, $\gamma_{\rm r}$ and $\gamma_{\rm nr}$ are the radiative and nonradiative contributions to the decay rate, and $s(t)$ is the exciting pulse. To model the effects of the pump, we added a time dependence to the properties of our mode ($\gamma_{\rm nr}$ and $\omega$). We use a $\sech^2$ excitation pulse: $s(t) = \sech^2((t)/\sigma_{\rm probe})\exp[-i\omega_{\rm probe}t]$, here $\sigma_{\rm probe}$ controlled the width of the pulse and $\omega_{\rm probe}$ determines the central frequency.

We determined the time dependence of the resonant mode by the following consideration. Since we used a short, Gaussian-type pump pulse to excite free carriers in the metasurface,  we can approximate the change of the dielectric permittivity of germanium by the error function: $\varepsilon(t) = \epsilon_{-\infty} + \Delta \varepsilon/2[\erf[(t+\tau)/\sigma_{\rm pump}]+1]$, where $\epsilon_{-\infty}$ is the permittivity without the pump, $\Delta \varepsilon$ is the total change induced in the material, $\tau$ is the pump arrival time, and $\sigma_{\rm pump}$ is the duration of the pump pulse. Assuming that the mode resonant frequency is linearly dependent on the permittivity, we can write the same equation for $\omega$ by substituting $\epsilon_{-\infty}$ for the unperturbed resonant frequency  $\omega_{-\infty}$ and $\Delta \varepsilon$ for the change in resonant frequency $\Delta \omega$:

\begin{equation}
\omega(t) = \omega_{-\infty} + \frac{\Delta \omega}{2}\left[\erf\left(\frac{t+\tau}{\sigma_{\rm pump}}\right)+1\right]
\label{"w(t)"}
\end{equation}

In addition to changing the resonant frequency, the free carriers also add absorption, which lead to a change in $\gamma_{\rm nr}$. Since permittivity and absorption generally have the same dependence on the free carrier concentration, here we assumed that $\gamma_{\rm nr}(t)$ behaved the same way $\omega(t)$ does: $\gamma_{\rm nr}(t) - \gamma_{\rm nr -\infty} = \xi(\omega(t)-\omega_{-\infty})$, here $\gamma_{nr -\infty}$ is the unperturbed nonradiative decay rate, and $\xi$ is a numeric coefficient that is determined by the material used for the resonator. Since the correct value of $\xi$ is unknown for aGe, we performed these simulations for various values from $\xi=0.5$ to $2$. We found that $\xi = 1$ yielded results closest to the experimental measurements, so we used this value for all the data presented here. Since the relative change of the refractive index of aGe is small, we assumed that the mode profile remains largely unchanged, and $\gamma_{\rm r}$ did not change significantly in the process, so we take it to be constant.

Equation \ref{"CMT"} was numerically solved using the explicit Runge-Kutta method as implemented in the NDSolve function from Wolfram Mathematica computation engine. The parameters we used were close to the experimental ones: $\omega_{-\infty} = 1150$ THz ($\lambda_{-\infty} = 1640$ nm), $\Delta \omega = 63$ THz ($\Delta \lambda = 100$ nm), $\sigma_{\rm probe} =\sigma_{\rm pump} = 50$ fs, $\omega_{\rm probe} = 1158$ THz($\lambda = 1628$ nm, for linear calculations) and $1171$ THz ($\lambda = 1610$ nm, for nonlinear calculations), $\gamma_{\rm nr, -\infty} = 4.5$ THz, $\gamma_{\rm r} = 4.5$ THz (equivalent to $Q=65$ for linear calculations) and $8$ THz (equivalent to $Q=36.6$ for nonlinear calculations). The two different decay rates and central wavelengths were used due to the different probe intensities and wavelengths used in the linear and nonlinear experiments. Due to difficulty in measuring the third harmonic spectra, a probe of higher intensity at $\lambda = 1610$ nm was used in nonlinear experiments, generating free carriers in germanium increasing the nonradiative losses. A much weaker probe at $\lambda = 1628$ nm is used in linear experiments, leading to no noticeable changes in the quality factor of the resonance. To model the pump--probe experiment, the pump arrival time $\tau$ from Eq. (\ref{"w(t)"}) is varied between $\tau = -1000$ fs and $\tau = 500$ fs. 

The light radiated by the mode can be obtained using the following equation: $s_{\rm r}(t) = s(t) - \sqrt{\gamma_{\rm r}}a(t)$. We then performed a Fourier transform: $I^{\rm simul}(\omega) = \mbox{abs}[F[s_{\rm r}(t)]]^2$, where $I^{\rm simul}(\omega)$ is the simulated transmission spectrum, and $F$ denotes the Fourier transform. The resulting transmission spectra as a function of the delay between the probe and the pump are shown in Fig.~\ref{fig:4}b. By performing the same calculation without any temporal changes to the resonant mode, we get the spectrum for the case when there is no pump present (the transmission in this case is shown with a black curve on Fig.~\ref{fig:4}a), then we can plot the same value as presented in experimental results $|I^{\rm simul}_{\rm pump} - I^{\rm simul}_{\rm no pump}|$ (Fig.~\ref{fig:4}c).

To simulate the third harmonic generation, the real part of the complex amplitude of the resonant mode was raised to the third power: $I^{THG}(\omega) = \mbox{abs}[F(Re[a(t)]^3)]^2$. Although this approach ignores several important effects that influence third harmonic generation in the system, it gives a good estimate of frequency conversion in the third harmonic generation spectrum. The normalized calculated nonlinear spectra are shown in Fig.~\ref{fig:4}d. The spectra for several pump--probe delays are shown on Fig.~\ref{fig:4}e on the logarithmic scale, where a blue-shift of the spectrum is clearly seen as the delay approaches zero. By fitting these spectra with a $\sech^2$ function, we estimate the width and central wavelength of the spectra, these results are shown on Fig.~\ref{fig:4}f,g. 

These simulation results provide excellent agreement with experimental data in both linear and the third harmonic spectra. We see the same periodic modulations in the linear spectra, that stretch far from the original probe spectrum (Fig.~\ref{fig:4}c). In third harmonic simulations we observe a shift in central frequency of $10$ nm and a widening of the spectrum by approximately 40\%, which is very close to the experimental results.

\begin{acknowledgement}

The work was supported by Russian Science Foundation (18-12-00475, optical measurements and coupled-mode theory) and Ministry of Science and Higher Education of Russian Federation (14.W03.31.0008, linear spectroscopy and FDTD simulation). The research was partly supported by MSU Quantum Technology Centre. The fabrication effort was carried out at the Cornell NanoScale Facility, a member of the National Nanotechnology Coordinated Infrastructure (NNCI), which is supported by the National Science Foundation (Grant NNCI-2025233).

\end{acknowledgement}

\begin{suppinfo}
CMT calculations of different resonance Q-factors.
\end{suppinfo}
}


\bibliography{main_Ge}

\end{document}


\subsection{Supplementary note 1: coupled-mode theory calculations for different Q-factors}
We used the coupled-mode theory (CMT)\cite{haus1984waves} (for details of calculations,  see Methods of the main text) to compare the cases of different resonance Q-factors. To achieve the Q-factor in the range of Q = $2 \div 300$, $\gamma_{NonRad -\infty}$ and $\gamma_{Rad}$ were varied. Other parameters were close to those in the experiment: $\omega_{-\infty} = 1150$ THz ($\lambda_{-\infty} = 1640$ nm), $\Delta \omega = 63$ THz ($\Delta \lambda = 100$ nm), $\sigma_{probe} =\sigma_{pump} = 50$ fs, $\omega_{probe} = 1171$ THz($\lambda = 1610$ nm). To model the pump--probe experiment, the pump arrival time $\tau$ from Eq.~3 from the main manuscript is varied between $\tau = -1000$ fs and $\tau = 500$~fs. We use a Gaussian excitation pulse: $s(t) = exp[-t^2/\sigma_{probe}^2]*exp[-i\omega_{probe} t]$, here $\sigma_{probe}$ is the duration of the pulse, and $\omega_{probe}$ is its central frequency.

\begin{figure}
\includegraphics[width=0.55\paperwidth]{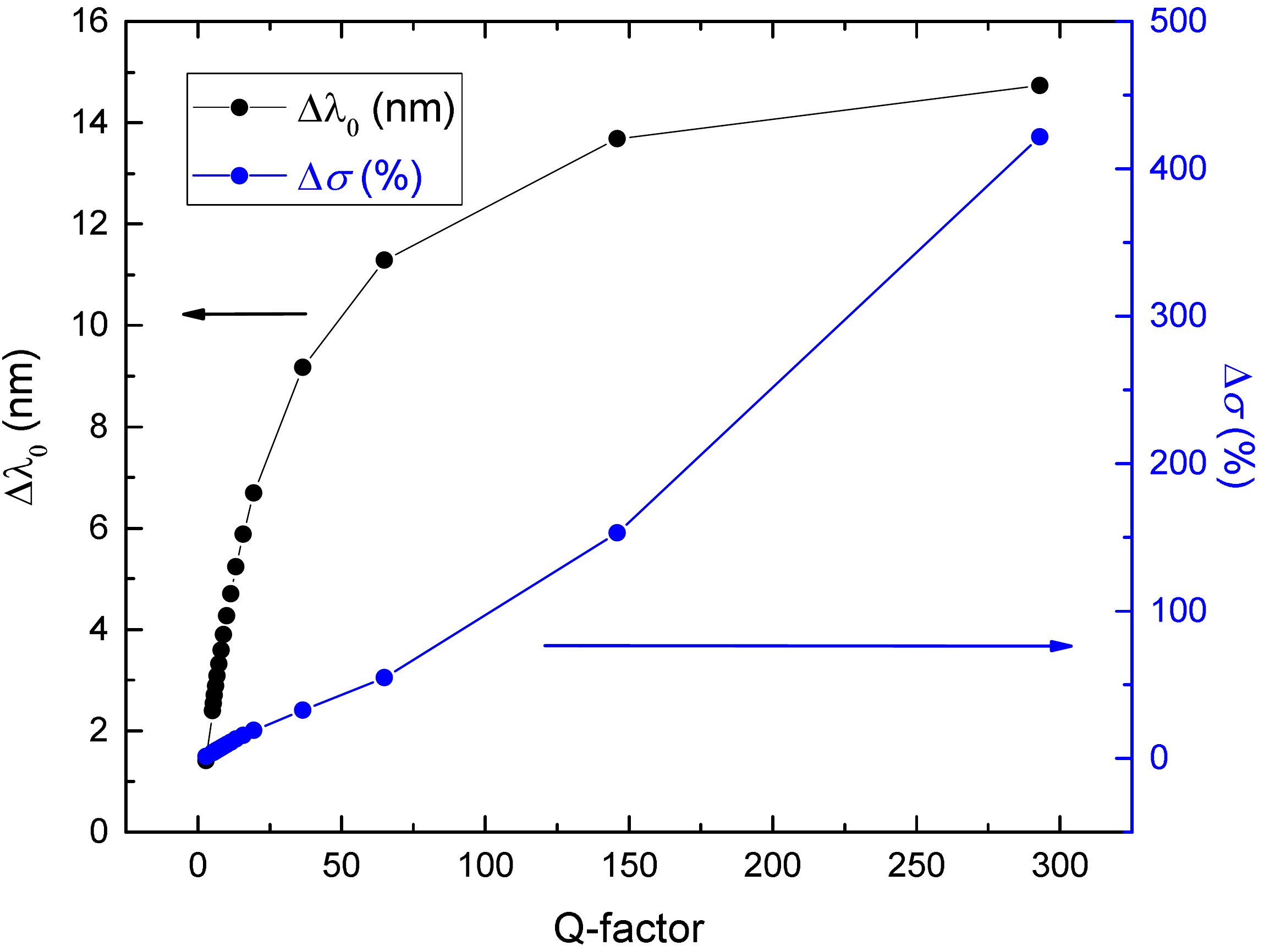}
\caption{\label{Qfactor} The fitting results for the spectral width $\sigma$ and central wavelength $\lambda_0$ of THG spectra by a Gaussian pulse with Eq.~S(\ref{gauss}) for  resonances with different Q-factors. $\Delta \lambda_0 = \lambda_0^{\tau=-1000\text{fs}} - \lambda_0^{\tau=-2.5\text{fs}}$ and $\Delta \sigma = \sigma^{\tau=-1000\text{fs}} - \sigma^{\tau=-2.5\text{fs}}$. }
\end{figure}

We fitted the calculated spectra of THG under pump illumination at time delay values $\tau = -1000$ and $\tau = -2.5$~fs by a Gaussian function with the following equation:

 \begin{equation}
y = y_0 + Ae^{-\frac{(\lambda-\lambda_0)^2}{2\sigma^2}},
\label{gauss}
\end{equation}
where $y_0$ is the offset , $A$~--- amplitude, $\lambda_0$~--- central wavelength and $\sigma$ is the width of the spectrum. The full width at half maximum (FWHM) is related to the width of the spectrum by the following relation FWHM$=2*\sigma\sqrt{2\ln{(2)}}$ for a Gaussian pulse.

Time delay value $\tau = -1000$ is similar to the case of the pump absence and $\tau = -2.5$~fs is time delay value near the most pronounced changes for both central wavelength of the spectra $x_c$ and spectral width $\sigma$. The values shown in Fig.~S1 are defied as follows: \\$\Delta \sigma = \sigma^{-1000\text{fs}} - \sigma^{-2.5\text{fs}}$ and $\Delta \lambda_0 = \lambda_0^{-1000\text{fs}} - \lambda_0^{-2.5\text{fs}}$ and demonstrate that higher Q-factor time-variant metasurfaces exhibit greater blue-shift and spectrum broadening of THG signal which can be detected in the experiment. Therefore, we have established the importance of the high Q-factor in the process of frequency conversion in time-variant metasurfaces. 

\bibliography{main_Ge}